\documentclass[12pt,aps,showpacs]{revtex4}
\usepackage{epsfig,amssymb,amsfonts,bm}
\makeatletter
\renewcommand{\@makefntext}[1]{\parindent=1em\noindent\hbox to
1.8em{\hss$^{\@thefnmark}$}#1}
\renewcommand{\@footnotemark}{\hbox{\mathsurround=0pt$^{\@thefnmark}$}}
\newcommand{\ftnote}[2]{\footnotemark[#1]\footnotetext[#1]{#2}}
\makeatother
\newcommand{\be} {\begin{equation}}
\newcommand{\ee} {\end{equation}}
\newcommand{\bdm} {\begin{displaymath}}
\newcommand{\edm} {\end{displaymath}}
\newcommand{\bc} {\begin{center}}
\newcommand{\ec} {\end{center}}
\newcommand{\beqa} {\begin{eqnarray}}
\newcommand{\eeqa} {\end{eqnarray}}
\newcommand{\nn} {\nonumber}

\newcommand{\bfig} {\begin{figure}}
\newcommand{\efig} {\end{figure}}
\newcommand{\btab} {\begin{tabular}}
\newcommand{\etab} {\end{tabular}}

\newcommand{\ep} {\epsilon}
\newcommand{\ga} {\gamma}
\DeclareMathSymbol{\varGamma}{\mathord}{letters}{"00}
\def\P{I\!\!P}
\def\R{I\!\!R}
\usepackage[usenames]{color}
\begin{document}

\title{Photoproduction of $a_0(980)$ and $f_0(980)$}

\author{A. Donnachie}
\affiliation{School of Physics and Astronomy, University of Manchester,
Manchester M13 9PL, England}

\author{Yu S Kalashnikova}
\affiliation{ITEP, 117259 Moscow, Russia}

\begin{abstract}
\noindent
There is no generally accepted view of the structure of the light-quark
non-strange scalar mesons. A variety of models has been proposed that
encompass $qq\bar{q}\bar{q}$, molecular, $q\bar{q}$ and glueball states
in various combinations. Previously we considered scalar-meson
photoproduction in a simple Regge-pole model and showed that it was
experimentally viable. Recent data on the photoproduction of $a_0(980)$
and $f_0(980)$ confirm this. We extend our model to incorporate Regge cuts,
based on our knowledge of $\pi^0$ photoproduction. The theoretical predictions
are compared to the $a_0(980)$ and $f_0(980)$ photoproduction data.
\end{abstract}
\pacs{13.60.Le, 12.39.Mk, 14.40.Be} \maketitle

\section{Introduction}
\label{intro}

The scalar sector of light--quark spectroscopy remains poorly understood,
and various phenomenological models have been suggested to describe the
light scalars. Simple ground or excited $^3P_0$ $q \bar q$ states
and tetraquark models have been discussed, as well as a glueball admixture
for the isoscalar--scalar states. Large branching ratios of scalars to
pseudoscalar meson pairs suggests also the possibility of scalar resonances
generated dynamically. The latter has been widely discussed in connection with
the $a_0(980)$ and $f_0(980)$ states which are very close to the $K \bar K$
threshold and, as such, could contain a large admixture of a $K \bar K$
molecule.

The photoproduction of $\pi \pi$/$\pi \eta$ and $K \bar K$ pairs near the
$K \bar K$ threshold is recognised as a powerful tool to study the properties
of the $f_0(980)$ and $a_0(980)$ mesons and to help
discriminate among models for scalars. If there is a large admixture of
pseudoscalar meson pairs, the resonance will be seen in the final state
interaction of the produced $\pi \pi$/$\pi \eta$ and $K \bar K$ pairs,
a mechanism that was considered in Refs. \cite{Sz98,Sz04}.
If, however, the scalar contains significant admixture of a compact
$q \bar q$ state, the resonance can be produced directly via the $q \bar q$
component. It is this possibility that is discussed in the present paper.
In reality the scalar meson wavefunction contains both $q \bar q$ and $K \bar K$ admixtures. We do not argue that the $q\bar{q}$ component is
the sole or even dominant part of the $a_0(980)$ and $f_0(980)$
wave function but do attempt to put reasonable limits on its
contribution to the photoproduction cross section. The calculation
does not exclude the possibility of a $K\bar{K}$ component in
either meson or indeed a glueball component in $f_0(980)$.

The photoproduction of the light-quark scalar mesons $a_0(980)$, $f_0(980)$,
$f_0(1370)$, $a_0(1450)$, $f_0(1500)$ and $f_0(1710)$ was calculated in
Ref. \cite{DK08}. The principal objective was to demonstrate that the cross
sections are sufficiently large to be measurable and that they provide a
viable mechanism to probe the structure of the scalar mesons. A simple model
was used, assuming the dominant mechanism to be Reggeised $\rho$ and $\omega$
exchange, both of which are well understood in pion photoproduction. The
electromagnetic couplings $\gamma S V$ of a vector meson to a scalar meson
were calculated assuming that the scalar mesons are bound $q \bar q$ $^3P_0$
states, so the radiative decays $V \to S\gamma$ and $S \to V\gamma$ proceed
via a quark loop and the corresponding matrix element can be estimated in the
quark model \cite{CDK02,CDK03}. These estimates were used in Ref. \cite{DK08} to
calculate the $a_0(980)$ and $f_0(980)$ photoproduction amplitude in the quark
loop mechanism and it was shown that even with a modest admixture of
$q \bar q$ component in the scalar wave function the direct mechanism will
dominate the cross section.

The latter result can be explained with the findings of
Ref. \cite{KKNHH06}: the radiative transitions between vector mesons and
$a_0(980)/f_0(980)$ exhibit a distinct hierarchy pattern so that
the closer is the mass of the vector meson to the $K \bar K$ threshold, the 
larger
is the transition via an intermediate kaon loop. For example, the
$\phi \to \gamma a_0(980)/f_0(980)$ amplitude is dominated by the kaon loop
mechanism, while $a_0(980)/f_0(980) \to \gamma \rho/\omega$ transition widths
are much smaller in the $K \bar K$ molecular model for scalars than in the
$q \bar q$ model. Clearly, photoproduction kinematics with $t$-channel vector
meson exchange discriminates even more strongly in favour of the quark loop
mechanism.

The conclusions of Ref. \cite{DK08} were that light-quark scalar meson
photoproduction on protons is a practical proposition given the luminosities
available to modern photoproduction experiments. However, contributions from
lower-lying trajectories, particularly that associated
with the $b_1(1235)$, and Regge cuts were not considered. The resulting
differential cross sections have a deep minimum in the vicinity of $t = -0.5$
GeV$^2$ due to the wrong-signature zeros in the $\rho$ and $\omega$
trajectories. Subsequently data on the photoproduction of $a_0(980)$
\cite{ELSA08} in the range $2.0 < E_\gamma < 2.85$ GeV and of $f_0(980)$
\cite{CLAS09} in the range $3.0 < E_\gamma < 3.8$ GeV have become available.
Neither cross section shows the minimum expected from the wrong-signature
zeroes in the $\rho$ and $\omega$ trajectories and both are larger than the
predictions of Ref. \cite{DK08} at small $t$. This is analogous to the situation
in $\pi^0$ photoproduction where strong cuts are required in both natural
and unnatural parity exchange \cite{BS78}. The analysis of Ref. \cite{BS78} made
use of finite energy sum rules (FESRs) and it was possible to make a clear
separation between the Regge-pole and Regge-cut contributions. The cuts do
not conform to any particular Regge-cut model and have to be treated
phenomenologically. A similar result holds in $\pi^+$ photoproduction
\cite{Wor72}. It is logical to assume that the discrepancy between the
results of Ref. \cite{DK08} and the data of Refs. \cite{ELSA08,CLAS09} for 
scalar photoproduction are
due primarily to the same kind of cut effects that occur in $\pi^0$
photoproduction.

It is not practical to analyse the data on
$a_0(980)$ and $f_0(980)$ photoproduction directly as both occur
at only one energy and with rather low statistics.  Thus we employ
the following strategy: as Regge-cut effects cannot be calculated
{\it a priori} we propose a simple phenomenological model for cut
effects that gives a reasonable description of $\pi^0$
photoproduction. We do not claim that this model of $\pi^0$ photoproduction
is on a par with more sophisticated approaches as, for example, that of
Ref. \cite{BS78}.
We use the existing data on
$\pi^0$ photoproduction to find the parameters of our simple model which
is readily transportable to scalar-meson
photoproduction.

A brief overview of Regge-cut phenomenology is given in Sec. II and is
applied to $\pi^0$ photoproduction in Sec. III. The model is extended to
$a_0(980)$ and $f_0(980)$ photoproduction in Sec. IV.
The implications of these results for exploring the nature
of the scalar mesons in photoproduction are discussed in
Sec. V.

\section{Regge-cut phenomenology}

The aim is to construct a simple model of $\pi^0$ photoproduction that
can be transported to scalar-meson photoproduction. This approach can be
justified as the Regge terms are the same in both cases: dominant natural
parity $\omega$ and $\rho$ exchange plus a small contribution from unnatural
parity $b_1(1235)$ exchange.
We know \cite{BS78} that cut
effects are important in $\pi^0$ photoproduction so we give a brief discussion
of the relevant phenomenology. A fuller discussion of Regge cuts may be
found in Ref. \cite{DDLN}.

Regge cuts arise from rescattering, the simplest being the exchange of two
Reggeons $R_1$ and $R_2$ although there is no reason to exclude multi-Reggeon
cuts. The exchange of two Reggeons with linear trajectories
$\alpha_i(t) = \alpha_i(0)+\alpha_i^\prime t$, $i$=1,2
are known \cite{LP72} to yield a cut
with a linear trajectory $\alpha_c(t)$:
\be
\alpha_c(t) = \alpha_c(0)+\alpha_c^\prime t
\label{cut1}
\ee
where
\beqa
\alpha_c(0) &=& \alpha_1(0)+\alpha_2(0)-1\nn\\
\alpha_c^\prime &=& \frac{\alpha_1^\prime\alpha_2^\prime}
{\alpha_1^\prime
+\alpha_2^\prime}.
\label{ctraj}
\eeqa

Cuts cannot be calculated with any precision and none of the numerous models
proposed agree with all aspects of the data. Hence cuts are best treated
phenomenologically, although even then there is no consistency among
different reactions. Good examples of this are provided by $\pi^- p \to
\pi^0 n$, dominated by $\rho$ exchange, and $\gamma p \to \pi^0 p$, dominated
by $\omega$ exchange. Effective $\rho$ and $\omega$ trajectories can be
obtained directly from the data: for example, see Figs. 5.1 and 5.7 in
Ref. \cite{CDG}. In the case of the $\rho$, the effective trajectory agrees rather
well with the extrapolation from the physical region, so cut effects are small
in $\pi^- p$ charge exchange. In the case of the $\omega$ there is essentially
no agreement with the extrapolation from the physical region, so cut effects
are significant. The FESR analysis of $\pi^0$ photoproduction by 
Ref. \cite{BS78} demonstrates this latter point explicitly.

As there is angular momentum associated with the two-Reggeon system, a Regge cut
will occur in natural and unnatural parity-exchange amplitudes irrespective
of the intrinsic parities of the two Reggeons. Naturality is defined as $+1$ if
the spin and parity of the mesons on it have natural parity and as $-1$ if they
have unnatural parity. It has been shown \cite{JL75} that if
the exchanged Reggeons have naturalities $n_1$ and $n_2$ then amplitudes of
naturality $-n_1n_2$ are suppressed relative to amplitudes of naturality
$+n_1 n_2$ and this suppression grows with increasing energy.

In the case of $\pi^0$ photoproduction one would expect the leading process
to be $\pi^0$ production followed by $\pi^0$ elastic scattering. The latter
is dominated by $f_2$ and Pomeron exchange, both of which have natural parity,
so the natural parity cut should dominate over the unnatural parity. This is
again in accord with the analysis of Ref. \cite{BS78}. Two cut terms are 
included in that analysis with trajectories $\alpha_3(t) = 0.447+0.333t$ and
$\alpha_4(t) = 0.177+0.5t$.

As in Ref. \cite{DK08} we assume linear non-degenerate $\omega$ and $\rho$
trajectories,
\beqa
\alpha_\omega &=& 0.44+0.9t\nn\\
\alpha_\rho &=& 0.55+0.8t.
\label{vtraj}
\eeqa
We take the Pomeron trajectory to be \cite{DDLN}
\be
\alpha_{\P} \sim 1.08+0.25t,
\label{pomtraj}
\ee

so from (\ref{ctraj}) the trajectory of the $\omega$-$\P$ cut is
$\alpha_{\P}^c = 0.52+0.196t$ and that of the $\rho$-$\P$ cut is
$\alpha_{\P}^c = 0.64+0.160t$. The second state on the $f_2$ trajectory is the
$f_4(2050)$ with a mass of 2018 MeV \cite{PDG}. Taking a mass of 1275 MeV for
the $f_2(1270)$ \cite{PDG} gives the $f_2$ trajectory as $\alpha_{f_2} =
0.672+0.817t$ and the trajectories of the associated $\omega$-$f_2$ and
$\rho$-$f_2$ cuts are $\alpha_{\R}^c = 0.112+0.428t$ and $\alpha_{\R}^c =
0.222+0.404t$. Thus it is reasonable to conclude that the $\alpha_3(t)$ cut
of Ref. \cite{BS78} corresponds roughly to the combined $\omega$-$\P$ and
$\rho$-$\P$ cuts and the $\alpha_4(t)$ cut corresponds roughly to the
combined $\omega$-$f_2$ and $\rho$-$f_2$ cuts.

\section{$\pi^0$ photoproduction}

\subsection{Vector exchange}

Let $q$, $p_1$, $k$, and $p_2$ be respectively the 4-momenta of the photon,
initial proton, pion and recoil proton. The hadronic current for vector
exchange is
\beqa
J^V_\mu&=&-e\frac{g_{V\pi\gamma}}{m_\pi}\ep_{\mu\nu\rho\sigma}
q_\nu p_\rho g_{\sigma\tau}\nn\\
&&\times\bar{u}(p_2)\big\{ig_V\gamma_\tau-g_T\sigma_{\tau\lambda}
p_\lambda\big\}u(p_1)D_V(s,t),
\label{vexch1}
\eeqa
where $m_\pi$ is the $\pi^0$ mass, $p = p_2-p_1$ and $D_V(s,t)$ is the full
Regge propagator for vector exchange:
\be
D_V(s,t)=
\Big(\frac{s}{s_0}\Big)^{\alpha_V(t)-1}
\frac{\pi\alpha^\prime_V}{\sin(\pi\alpha_V(t))}
\frac{-1+e^{-i\pi\alpha_V(t)}}{2}\frac{1}{\Gamma(\alpha_V(t))}.
\label{regge}
\ee
with $\alpha_V(t) = \alpha_{V0}+\alpha_V^\prime t$ the Regge trajectory.
As in Ref. \cite{DK08} we use $g_V^\omega = 15$, $g_T^\omega = 0$,
$g_V^\rho = 3.4$ and $g_T^\rho = 11$ GeV$^{-1}$.
The electromagnetic coupling
constants $g_{V\pi\gamma}$ are obtained from the electromagnetic
decay width:
\be
\Gamma_{(V \to \pi\gamma)} = \frac{\alpha}{24}\Big\{\frac{g_{V\pi\gamma}}
{m_\pi}\Big\}^2m_V^3\Big\{1-\Big(\frac{m_\pi}{m_V}\Big)^2\Big\}^3
\label{vpigam}.
\ee
This gives $g_{\omega\pi\gamma} = 0.322$ for a width of 75.6 keV \cite{PDG}
and $g_{\rho^0\pi\gamma} = 0.119$ for a width of 89.6 keV \cite{PDG}.

Note that the vector-exchange contributions vanish at the wrong-signature
points given by $\alpha_V(t) = 0$ i.e. at $t = -0.49$ GeV$^2$ for the $\omega$
and $t = -0.69$ GeV$^2$ for the $\rho$. These zeros result in a pronounced
dip in the differential cross section.

The cross section for the exchange of a single vector meson is

\be
\frac{d\sigma}{dt} = -\frac{T_V}{16\pi(s-m_p^2)^2}
\label{vsig1}
\ee

where
\beqa
T_V &=& \frac{4\pi\alpha  g_{V\pi\gamma}^2}{m_\pi^2}\{\frac{1}{2}
[s(t-t_1)(t-t_2)+\frac{1}{2}t(t-m_{\pi}^2)^2]aa^*\nn\\
&&+\frac{1}{2}m_ps(t-t_1)(t-t_2)(ab^*+a^*b)\nn\\
&&+\frac{1}{8}s(4m_p^2-t)(t-t_1)(t-t_2)bb^*\}|D_V(s,t)|^2.
\label{vsig2}
\eeqa

Here $t_1$ and $t_2$ are the kinematical boundaries,
\beqa
t_{1,2} &=& \frac{1}{2s}\{-(m_p^2-s)^2+m_{\pi}^2(m_p^2+s)\nn\\
&& \pm(m_p^2-s)\sqrt{(m_p^2-s)^2-2m_{\pi}^2(m_p^2+s)+m_{\pi}^4}\},
\eeqa
and
\be
a = (g_V+2m_p g_T)~~~~~~b = -2g_T
\label{abv}
\ee

\subsection{Axial-vector exchange}

It is possible to separate natural-parity and unnatural-parity exchange in pion
photoproduction by using a plane-polarised photon beam. The polarised-beam
asymmetry $(\sigma_\perp-\sigma_\parallel)/(\sigma_\perp+\sigma_\parallel)$
is close to unity for natural-parity exchange and any deviation from
this indicates the presence of unnatural-parity exchange. This is clearly
the case in $\pi^0$ photoproduction as can be seen in Figs. 1 and 2 below.

The Regge-pole exchange is that associated with the $b_1(1235)$. $C$-parity
requires that the coupling of the $b_1(1235)$ to the nucleon is the
axial-tensor coupling $\sigma_{\mu\nu}\gamma_5$ \cite{KM10}. The hadronic
current for $b_1$ exchange may be written as
\be
J^b_\mu = -e g_b g_{b_1NN}\big\{(p.q)g_{\mu\beta}-
p_\mu q_\beta \big\}(p_{2\beta}+p_{1\beta}) \bar{u}(p_2)
\gamma_5 u(p_1)D_b(s,t),
\label{bexch}
\ee
 where
\be
g_b = \frac{g_{b_1\pi\gamma}}{m_\pi}.
\label{gbdef}
\ee

The axial-vector Regge propagator $D_b(s,t)$ has the same form as
(\ref{regge}) with $\alpha_V(t)$ replaced by the $b_1$ trajectory
$\alpha_b(t)\approx -0.013+0.664t$.
Note that there is no interference between vector and
axial-vector exchange in the cross section or polarised-beam
asymmetry.

The cross section is
\be
\frac{d\sigma}{dt} = -\frac{T_A}{16\pi(s-m_p^2)^2}
\label{asig1}
\ee
where
\be
T_A = -4\pi\alpha\frac{g^2_{b_1\pi\gamma}}{m^2_\pi}g_{b_1NN}^2\frac{st}{2}
(t-t_1)(t-t_2)|D_b(s,t)|^2.
\label{asig2}
\ee

The value of $g_{b_1\pi\gamma}$ can be found from the radiative decay
width $\Gamma_{b_{1}^+\pi^+\gamma}$ which
is given by

\be
\Gamma_{b_1^+\pi^+\gamma} = \frac{\alpha}{24}
\Big\{\frac{g_{b_1\pi\gamma}}{m_\pi}\Big\}^2 m_{b_1}^3
\Big\{1-\Big(\frac{m_\pi}{m_{b_1}}\Big)^2\Big\}^3.
\label{apigam}
\ee

The radiative decay width of $b_1^+ \to \pi^+\gamma$ is
$\Gamma_{b_1^+\pi^+\gamma} = 228 \pm 57$ keV \cite{PDG}, so
$g_{b_1\pi\gamma}/m_\pi = 0.648 \pm 0.081$ GeV$^{-1}$. Although the
value of $g_b$ is rather well established, little is known about $g_{b_1NN}$.

We adopt the suggestion of Ref. \cite{KM10} that $g_{b_1NN}=G_{b_1NN}/2m_p$
with $G_{b_1NN} \approx G_{a_1NN}\approx 7$.

The contribution of $b_1(1235)$ exchange is negligibly small
for this choice of coupling and indeed remains small for any reasonable
value of $g_{b_1NN}$. Thus the unnatural-parity-exchange contribution must
come primarily from the unnatural-parity contribution arising from the
$\omega$ and $\rho$ cuts.

\subsection{The cut contributions}

As a physical mass cannot be be associated with a cut, the simplest form of
amplitude for a cut term is
\be
A_c(s,t) = C_c D_c(s,t)
\label{cutamp1}
\ee
where $C_c$ is a constant and
\be
D_c(s,t) = e^{d_ct}e^{-i\frac{1}{2}\pi\alpha_c(t)}s^{\alpha_c(t)-1}.
\label{cutamp2}
\ee
where we have retained only the Regge phase and absorbed
the rest of the $t$-dependence in the exponential, $\alpha_c(t)$ is the
cut trajectory and the constants $C_c$ and $d_c$ for each cut term are
obtained by fitting data.

We need a mechanism to allow us to transfer the $\pi^0$ cut model to scalar
photoproduction. The simplest way is to take the cut terms proportional to
the dominant $\omega$ and $\rho$ exchanges, retaining the kinematical
structure and replacing $g_{V\pi\gamma}g_{VNN}D_V(s,t)$, $V = \omega, \rho$
by
\be
g_{V\pi\gamma}g_{VNN}(D_V(s,t)+C^V_{n_1}D^V_{c_1}(s,t)
+C^V_{n_2}D^V_{c_2}(s,t)),
\label{ncut}
\ee
where $g_{V\pi\gamma}$ and $g_{VNN}$ are respectively the $V\pi\gamma$ and
relevant $VNN$ coupling constants and $C^V_{n_1}$ and $C^V_{n_2}$ are
respectively the natural-parity constants for the $V$-$f_2$ and $V$-$\P$ cuts.
These cuts also feed into the unnatural-parity exchange term and are much
larger than any cuts generated by $b_1(1235)$ exchange due to its small
contribution. So $g_bg_{B_1NN}D_b(s,t)$ is replaced by
\be
g_{b}g_{b_1NN}D_b(s,t) + \sum_V g_{V\pi\gamma}g_{VNN}
(C^V_{u_1}D_{c_1}(s,t)+C^V_{u_2}D_{c_2}(s,t)),
\label{ucut}
\ee
where the $C^V_{u_i}$ are the unnatural-parity constants.
It turns out that the cuts dominate unnatural parity exchange
so in practice the $b_1$ pole term could be omitted.

The parameters for $\rho$ and $\omega$ were taken to be the same i.e.
$C^\rho_{n_i} = C^\omega_{n_i}$ and $C^\rho_{u_i} = C^\omega_{u_i}$,
$i = 1,2$. Also the argument $d_c$ of the exponential in (\ref{cutamp2})
was taken to be the same for all terms. So in practice we have only five
free parameters to describe $\pi^0$ photoproduction.

This approach is plausible and has the merit of simplicity, although obviously
it is not unique. However as the aim is to provide a reasonable qualitative
description of $\pi^0$ photoproduction rather than a precise fit it is
perfectly adequate and, as is shown in the next section, is surprisingly
good.

\begin{center}
\begin{figure}
\epsfig{file=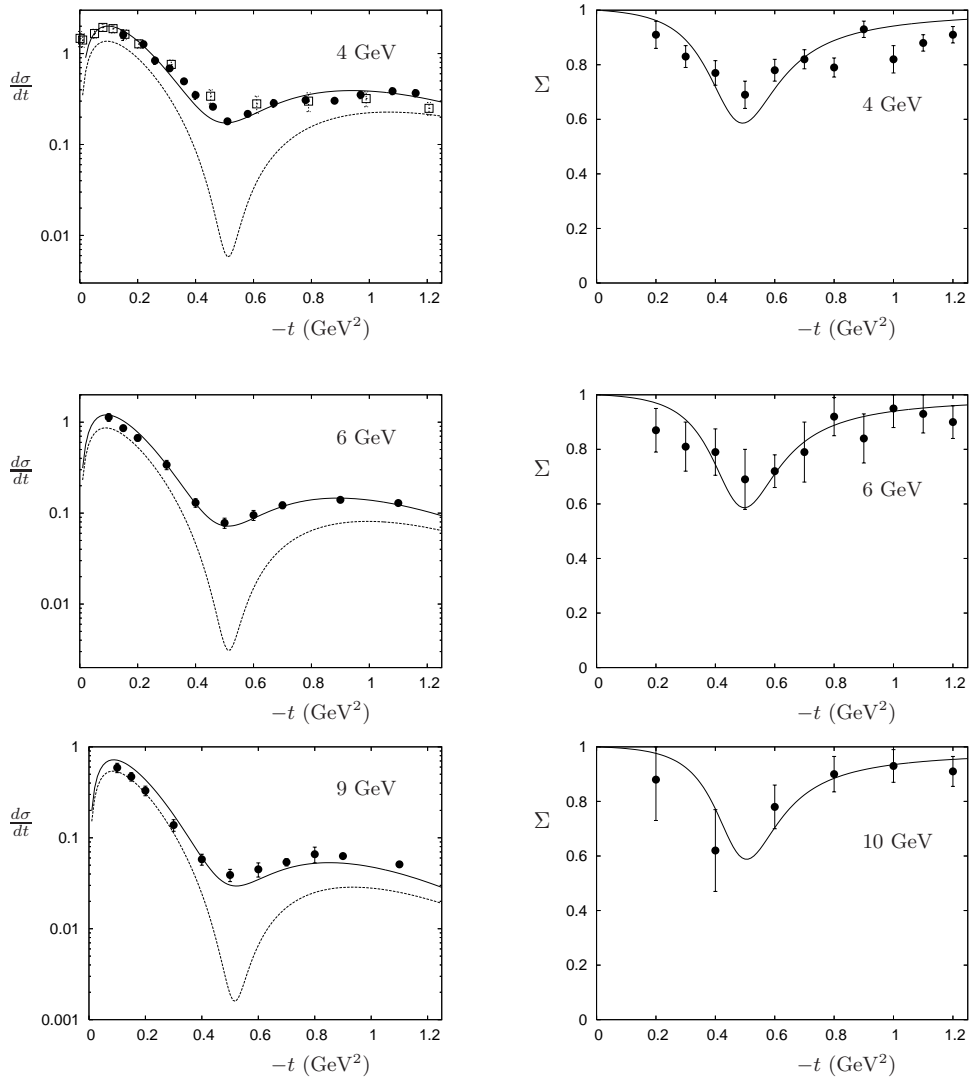, width=14cm}
\caption{Fits to the differential $\pi^0$ photoproduction cross sections
in $\mu$bGeV$^{-2}$ at $E_\gamma =$ 4, 6 and 9 GeV and the polarized-beam
asymmetry $\Sigma$ at $E_\gamma =$ 4, 6 and 10 GeV. Predictions without cuts
are shown as a dotted line. The differential cross section data at 4 GeV are
from Ref. \cite{Hufton73} (squares) and Ref. \cite{Brau68} (dots). The 
differential cross section data at 6 and 9 GeV and the polarized-beam 
asymmetry data are from Ref. \cite{And71}.}
\label{dsigdt1}
\end{figure}
\end{center}

\subsection{Fits to $\pi^0$ photoproduction}

For the fit we use the differential cross section data of the Liverpool group
at $E_\gamma$ = 4 GeV \cite{Hufton73}, of Braunschweig {\it et al} \cite{Brau68}
at $E_\gamma$ = 4, 5 and 5.8 GeV and Anderson {\it et al} \cite{And71} at
$E_\gamma$ = 6, 9 and 12 GeV. The polarized-beam asymmetry data are from
Anderson {\it et al} \cite{And71} at $E_\gamma$ = 4, 6 and 10 GeV. A typical
fit to some of the differential cross section data and the polarized-beam
asymmetry is shown in Fig. 1, which also gives the result
for the cross section without the cut terms.

Figure 2 compares the predictions of the model with differential cross section
data at $E_\gamma$ = 2.0 GeV \cite{Brau68,Dugger} and 3.0 GeV \cite{Brau68}
and polarized-beam data at $E_\gamma$ = 2.0 GeV \cite{Bussey75} and 3.0 GeV
\cite{Bell69}. Although these energies are rather low for the model,
particularly $E_\gamma$ = 2.0 GeV, the $a_0(980)$ and $f_0(980)$
photoproduction data are close to these energies. Nonetheless, the
extrapolation of the model to these low energies is sufficiently satisfactory
for our present purpose. The forward peak is reasonably well reproduced,
although rather low at 2 GeV, but the model retains some dip structure that
is not apparent in the data.

The contribution from unnatural-parity exchange is small, except in the dip
region. Elsewhere, as there is no interference between natural-parity and
unnatural-parity exchange, natural-parity exchange dominates. This is
immediately apparent from the polarized-beam asymmetry.

\begin{center}
\begin{figure}
\epsfig{file=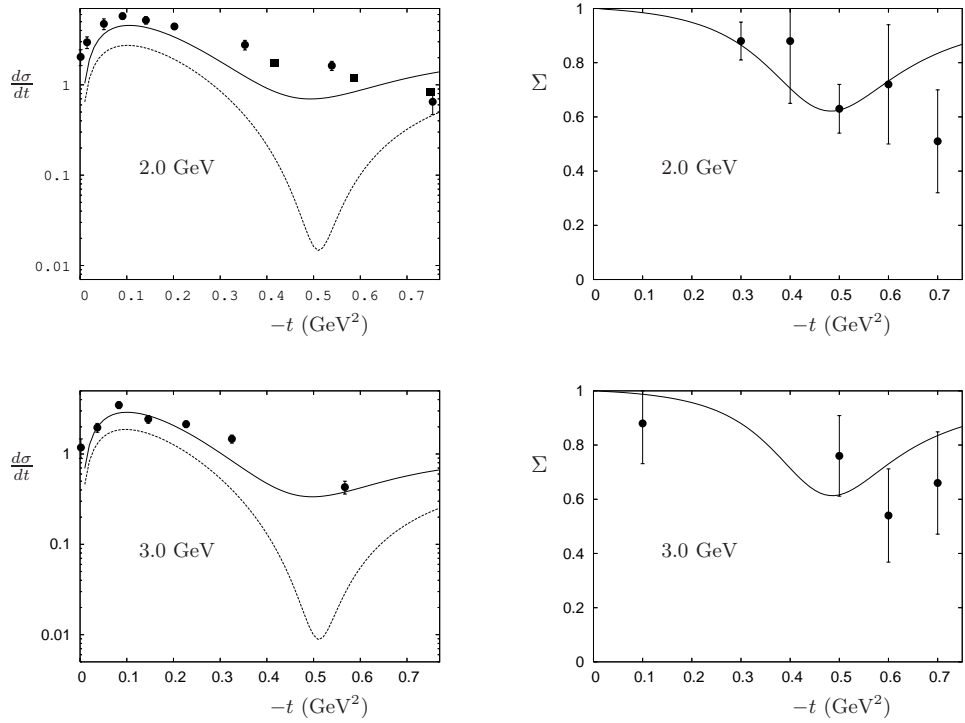, width=14cm}
\caption{Comparison of the model with the differential $\pi^0$ photoproduction
cross sections in $\mu$bGeV$^{-2}$ and with the polarized-beam asymmetry
$\Sigma$ at $E_\gamma =$ 2 and 3 GeV. Predictions without cuts are shown as the
dotted line. The differential cross section data at 2 GeV are from
Ref. \cite{Brau68} (dots) and Ref. \cite{Dugger} (squares) and at 3 GeV from
Ref. \cite{Brau68}. The polarized-beam data at 2 GeV are from 
Ref. \cite{Bussey75} and at 3 GeV from Ref. \cite{Bell69}.}
\label{dsigdt2}
\end{figure}
\end{center}

\section{$a_0(980)$ and $f_0(980)$ photoproduction}

\subsection{Vector exchange}

The hadronic current for vector exchange is \cite{DK08}
\be
J_\mu^V = \{g_{\mu\nu}(p.q)-p_\mu q_\nu\}
\bar{u}(p_2)\{a\ga_\nu+bp_{1\nu}\}u(p_1)D_V(s,t)
\ee
where $D_V(s,t)$ is the Regge propagator (\ref{regge}), $a = g_S(g_V+2m_pg_T)$,
$b = -2g_Sg_T$, with $g_V$ and $g_T$ as before, and $g_S$ is the coupling at
the $S V \ga$ vertex, defined in terms of the radiative decay width by
\be
\Gamma(S \to V\gamma) = g_S^2\frac{m_S^3}{32\pi}\Big(1-\frac{m_V^2}{m_S^2}
\Big)^3.
\ee
For the radiative decay width we use the results of Ref. \cite{DK08} which were
based on the model of Refs. \cite{CDK02,CDK03}. It is assumed that the scalar 
mesons
are $q\bar{q}$ $^3P_0$ bound states with the radiative decay proceeding via a
quark loop. Specifically
\beqa
\Gamma(a_0(980) \to \ga\rho) &=& 14~ {\rm keV}\nn\\
\Gamma(f_0(980) \to \ga\rho) &=& 125~ {\rm keV}.
\label{widths}
\eeqa
The radiative widths to $\ga\omega$ are a factor of 9 larger for the
$a_0(980)$ and a factor of 9 smaller for the $f_0(980)$.

It was shown in Refs. \cite{CDK02,CDK03} that the model gives good agreement 
with existing data on radiative decays of $q \bar q$ mesons.

The cross section is \cite{DK08}
\be
\frac{d\sigma}{dt} = -\frac{\tilde{T}_V}{16\pi(s-m_p^2)^2}
\label{vsigs1}
\ee
where
\beqa
\tilde{T}_V &=& {\textstyle\frac{1}{8}}g_S^2\{4aa^*[s(t-t_1)(t-t_2)+
2t(t-m_S^2)^2]\nn\\
&&+4(ab^*+a^*b)m_ps(t-t_1)(t-t_2)\nn\\
&&+bb^*s(4m_p^2-t)(t-t_1)(t-t_2)\}|D_V(s,t)|^2,
\label{vsigs2}
\eeqa
$a$ and $b$ are given by (\ref{abv}) and $t_1$ and $t_2$ are the
kinematical boundaries.  \ftnote{1}{It was pointed out in Ref. \cite{dSM12}
that there is a typographical error in equations (17) and (A.8) of
Ref. \cite{DK08}. The results presented in \cite{DK08} used the correct formula,
as given here.}

\subsection{Axial-vector exchange}

The structure of the current for $b_1$ exchange is
\be
J_\mu^A = -{\tilde g}_b\ep_{\mu\nu\lambda\rho}p_\lambda q_\rho
(p_{1\nu}+p_{2\nu})\bar{u}(p_2)\ga_5 u(p_1)D_b(s,t)
\ee
where $D_b(s,t)$ is the Regge propagator, and ${\tilde g}_b$ contains
$b_1S\gamma$ and $b_1NN$ couplings. The cross section is

\be
\frac{d\sigma}{dt} = -\frac{\tilde{T}_A}{16\pi(s-m_p^2)^2}
\label{bsigs1}
\ee
where
\be
\tilde{T}_A = -{\textstyle\frac{1}{2}}{\tilde g}_b^2
st(t-t_1)(t-t_2)|D_b(s,t)|^2.
\label{bsigs2}
\ee

As we know even less about the $b_1(1235)$ couplings in
scalar-meson photoproduction than in $\pi^0$ photoproduction the pole term
was again omitted, Eqs. (\ref{bsigs1}) and (\ref{bsigs2}) providing the
kinematical structure for the cut terms.
As in the case of $\pi^0$ photoproduction there is no interference
between natural-parity and unnatural-parity exchange
in the cross section.

\subsection{The cut contributions}

We adopt exactly the same approach for the cut contributions as for $\pi^0$
photoproduction, with the constants $C^V_{n_i}$ and $C^V_{u_i}$ having the
same values as in $\pi^0$ photoproduction. That is the relative strengths of
the cut and pole terms are the same as in $\pi^0$ photoproduction and the
structure of the cross section is the same as for scalar photoproduction
without the cut terms.

\subsection{Mass distributions}

To obtain mass distributions for $a_0(980)$ and $f_0(980)$ we represent
them as Breit-Wigner resonances with energy-dependent partial widths. The
signal cross section for the final state $i$ is given by \cite{DK08}
\be
\frac{d^2\sigma}{dt~dM} = \frac{d\sigma_0(t,M)}{dt}\frac{2m_S^2}{\pi}
\frac{\Gamma_i(M)}{(m_S^2-M^2)^2+M^2\Gamma_{\rm Tot}^2}
\label{bw0}
\ee
where $d\sigma_0(t,M)/dt$ is the narrow-width differential cross section
at a scalar mass M which is straightforward to calculate in our model. In
practice the narrow-width cross section varies very little over the width of
the scalar meson so it can be evaluated at the scalar mass and weighted with the
integral over the Breit-Wigner line shape. However, a problem arises in the
choice of the Breit-Wigner amplitude.

The Breit-Wigner amplitudes that have been used to describe the decays
$a_0(980) \to \pi^0\eta, K\bar{K}$ and $f_0(980) \to \pi\pi, K\bar{K}$
are those employed  by the KLOE Collaboration to analyse their data on $\phi \to
\pi\eta\gamma$ \cite{kloe09} and $\phi \to \pi^0\pi^0\gamma$ \cite{kloe07f},
and are based either on a ``kaon-Loop'' (KL) model \cite{achasov06} in which the
radiative transition proceeds via kaon loop, or on a ``no-structure'' (NS)
model
\cite{isidori06} in which the coupling is point-like. In the case of
$a_0(980)$ the $\pi\eta$ line shape in the KL model is
\be
B_a=t_{\pi\eta}^at_{\pi\eta}^{a*}\frac{k_{\pi\eta}}{4\pi^2},
\label{bw1}
\ee
where
\be
k_{\pi\eta}=\sqrt{\frac{(M^2-(m_{\pi}+m_{\eta})^2)(M^2-(m_{\pi}-m_{\eta})^2)}
{4M^2}}
\label{bw2}
\ee
 and
\be
t_{\pi\eta}^a=\frac{g_{\pi\eta}}{D_a},
\label{bw3}
\ee
$$
D_a=m_S^2-M^2+\mbox{Re}\Pi_{\pi\eta}(m_S)-\Pi_{\pi\eta}(M)+
$$
\be
\mbox{Re}\Pi_{K^+}(m_S)-\Pi_{K^+}(M)
+\mbox{Re}\Pi_{K^0}(m_S)-\Pi_{K^0}(M),
\label{bw4}
\ee
\be
\Pi_{\pi\eta}(M)=
\frac{g_{\pi\eta}^2}{16\pi}
\Big(\frac{m_+m_-}{\pi M^2}\log{\frac{m_{\pi}}{m_{\eta}}}+
\rho_{\pi\eta}(M)\Big(i+
\frac{1}{\pi}\log{\frac{1-\rho_{\pi\eta}(M)}{1+
\rho_{\pi\eta}(M)}}\Big)\Big),
\ee
\be
m_{\pm}=m_{\eta} \pm m_{\pi},
\ee
\be
\rho_{\pi\eta}(M)=\sqrt{(1 - m_+^2/M^2)(1 - m_-^2/M^2)},
\ee
$$
\Pi_{K}(M)=
\frac{1}{2}\theta(M-2m_K)
\frac{g_{K}^2}{16\pi}
\rho_{K}(M)\Big(i +
 \frac{1}{\pi}\log{\frac{1 - \rho_{K}(M)}{1 + \rho_{K}(M)}}\Big)
$$
\be
-\frac{1}{2}\theta(2m_K-M)\frac{g_{K}^2}{16\pi^2}
|\rho_{K}(M)|(\pi-2\arctan{|\rho_{K}(M)|}),
\ee
\be
\rho_K=\sqrt{1-\frac{4m_K^2}{M^2}}.
\ee
Here $g_K=\sqrt{2}g_{K^+K^-}$.

The NS line shape is given by the same expressions (\ref{bw1})
and (\ref{bw2}) with the replacement $D_a \to D_m$ where,
\be
D_m=m_S^2-M^2-\Sigma,
\ee
$$
\Sigma=i\frac{g_{\pi\eta}^2}{16\pi}
\rho_{\pi\eta}(M)+
$$
$$
\frac{i}{2}\theta(M-2m_{K^+})
\frac{g_{K}^2}{16\pi}
\rho_{K^+}(M)-\frac{1}{2}\theta(2m_K^+-M)\frac{g_{K}^2}{16\pi}
|\rho_{K^+}(M)|+
$$
\be
\frac{i}{2}\theta(M-2m_{K^0})
\frac{g_{K}^2}{16\pi}
\rho_{K^0}(M)-\frac{1}{2}\theta(2m_K^0-M)\frac{g_{K}^2}{16\pi}
|\rho_{K^0}(M)|.
\ee
The corresponding expressions for $f_0 \to \pi\pi$ are the same with the
obvious replacement of $m_\eta$ with $m_\pi$ and $g_{\pi\eta}$ with
$g_\pi = \sqrt{3/2}g_{\pi^+\pi^-}$.

Both the NS and KL forms were used by KLOE to analyse their data on $\phi \to
\pi\eta\gamma$ \cite{kloe09} and $\phi \to \pi^0\pi^0\gamma$ \cite{kloe07f},
updated in Ref. \cite{kloe08}, giving two parameter sets for each of $a_0(980)$
and $f_0(980)$. The KL formalism has also been applied to the reactions
$\gamma\gamma \to \eta\pi$ \cite{achasov10} to give two somewhat different
results for the $a_0(980)$, and to $\gamma\gamma \to \pi\pi$, together with
the $I = 0$ $S-$wave phase shift \cite{achasov11,achasov12}, giving 
eight further parameter sets for the $f_0(980)$.

Formally, there is no contradiction between the  KL and NS forms of
Breit-Wigner amplitudes and it can be easily verified that
the KL form reduces to
the NS one near the $K \bar{K}$ threshold. However there is a
significant difference when they are applied
to the production mechanism. In the key reactions $\phi \to \gamma S$
the transition mechanism is via a quark loop in the case of the NS
model {\it versus} a kaon loop in the case of the KL model. In other words, in
the absence of defined couplings, the model assumed for the transition
mechanism automatically leads to a difference in the $g_K$ coupling, it
being significantly larger in the KL case.

Additional information is available on the ratio $g_{K}^2/g_{\pi}^2$ for the
decay of the $f_0(980)$, principally from the analysis of $\pi\pi \to \pi\pi$
and $\pi\pi \to K\bar{K}$ \cite{martin77,estabrooks79,bugg96,kaminski99}, and
from $J/\psi \to \phi(1020)\pi^+\pi^-$ and $J/\psi \to \phi(1020)K^+K^-$
\cite{bes05}. The results from these different analyses are
consistent  and average to $g_{K}^2/g_{\pi}^2 = 4.0 \pm 0.3$ \cite{ochs13}.
None of these experiments quotes a width, so the actual values
of $g_{\pi}$ and $g_{K}$ are unknown. Consequently we do not use these results
for predictions in our model.

None of the ten parameter sets for the $f_0(980)$ from
Refs. \cite{kloe08,achasov11,achasov12} satisfies this constraint explicitly, but
nine have $g_{K}^2/g_{\pi}^2 > 4.0$ and one has $g_K^2 \ll g_\pi^2$. For
definiteness in the KL model we take \cite{achasov12}.
\be
m_{f_0} = 0.9783~{\rm GeV}~~~~g_K=5.006~~~~g_{\pi}=1.705,
\label{f0KL}
\ee
which is the one with the lowest $g_{K}^2/g_{\pi}^2$ value.
The NS model yields
$g_{K}^2/g_{\pi}^2 < 1$, and, to illustrate the sensitivity of our
results to the ratio of couplings,
we include it with the parameters \cite{kloe08}
\be
m_{f_0} = 0.9847~{\rm GeV}~~~~g_K=0.556~~~~g_{\pi}=1.600.
\label{f0NS}
\ee

There is no additional information on the $a_0(980)$ couplings, so for
consistency we again take the KL model set with the lowest
$g_{K}^2/g_{\pi\eta}^2$ value. This is \cite{kloe09}
\be
m_{a_0} = 0.9825~{\rm GeV}~~~~g_K=3.05~~~~g_{\pi\eta}=2.82.
\label{a0KL}
\ee
The corresponding result for the NS model is \cite{kloe09}
\be
m_{a_0} = 0.9825~{\rm GeV}~~~~g_K=2.22~~~~g_{\pi\eta}=2.16.
\label{a0NS}
\ee

\subsection{Prediction of $a_0(980)$ and $f_0(980)$ photoproduction}

In principle there are no free parameters in this calculation, as the couplings
are known and the constants defining the cut terms are determined by the fit to
$\pi^0$ photoproduction. However, the range of values for the constants for 
$a_0(980)$ and $f_0(980)$, particularly the latter, is such that a unique
prediction is not possible.

In Ref. \cite{DK08} three scenarios for the scalars were considered in which the
lowest $n\bar{n}$ nonet contains the $a_0(980)$ and $f_0(980)$ as members. In
two of these the $f_0(980)$ is mixed with either the $f_0(1370)$ or the
$f_0(1500)$ such that they are octets and the $f_0(980)$ is the singlet
$(u\bar{u}+d\bar{d}+s\bar{s})/\sqrt{3}$. In the third the $f_0(980)$ is the
standard $(u\bar{u}+d\bar{d})\sqrt{2}$ member of the octet. For definiteness
we use the latter.

\begin{center}
\begin{figure}
\epsfig{file=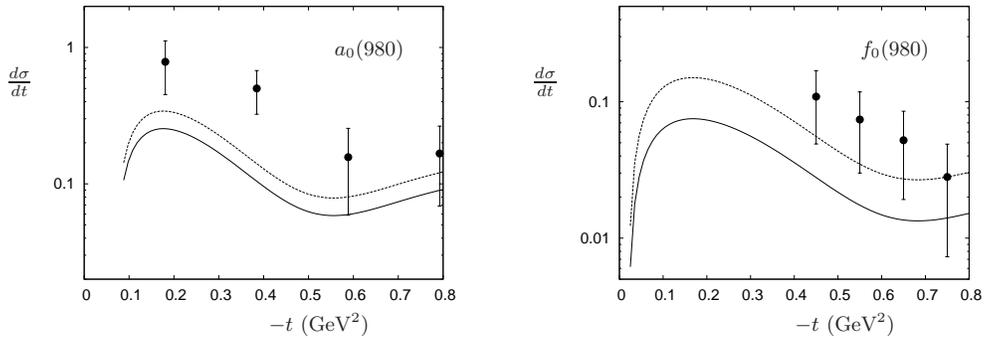, width=14cm}
\caption{Predictions of the differential cross sections for $a_0(980)$
photoproduction in the $\pi^0\eta^0$ channel (left) and $f_0(980)$ 
photoproduction in the $\pi^+\pi^-$ channel (right). The data are from 
the ELSA \cite{ELSA08}
and CLAS\cite{CLAS09} experiments respectively and the
units are $\mu b$ GeV$^{-2}$. In each plot the upper line is
the NS model prediction and the lower line is the KL model prediction}
\label{a0_f0_dsig}
\end{figure}
\end{center}

The $a_0(980)$ and $f_0(980)$ Breit-Wigner line shapes were integrated over the relevant
experimental ranges and the resulting cross sections for the KL
and NS
model parameters of (\ref{a0KL}), (\ref{a0NS})  and (\ref{f0KL}),
(\ref{f0NS})
are given in
Fig. \ref{a0_f0_dsig}, together with the ELSA \cite{ELSA08} and CLAS
\cite{CLAS09} small-$t$ data. In both cases the NS prediction is the upper
curve.

There are several reasons why one expects the predicted cross sections to be
smaller than the data. There is a coherent continuum background in the
$\pi^0\eta^0$ and $\pi^+\pi^-$ channels that will interfere with direct
production of the $a_0(980)$ and $f_0(980)$ respectively. The effect of this
is to increase the cross section and distort the resonance line shape.
This is discussed in detail in Ref. \cite{DK08}. Furthermore, there is 
the possibility
of background from the decay of high-mass baryon resonances. This is explicit
in the case of the $a_0(980)$ from $P_{33}(1232)\eta$ and $S_{11}(1535)\pi$:
see Fig. 4(i) of Ref. \cite{CLAS09}. Sideband subtraction certainly removes
incoherent background but not coherent background.
Finally we note that our results are sensitive to the choice
of the rescattering terms, particularly in natural-parity exchange through
interference with the leading terms.

The results show that the principal objective of Ref. \cite{DK08}, namely to
demonstrate that scalar photoproduction cross sections are sufficiently
large to be measured, has been attained. They also show that the production
mechanism is more complicated than that considered in Ref. \cite{DK08}, in
particular cut effects are not negligible. This is not particularly
surprising, but it does make more difficult the extraction of radiative
widths from photoproduction data. Within the context of the present model,
the pole and cut terms can be separated via the energy dependence of the
cross section.

\begin{center}
\begin{figure}
\epsfig{file=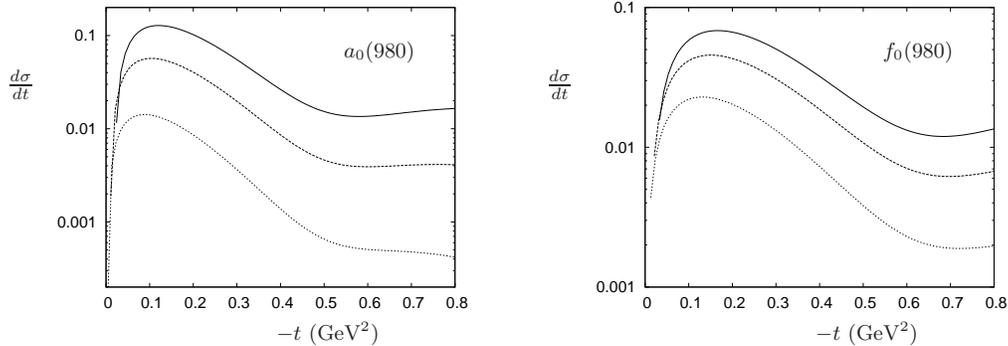, width=14cm}
\caption{NS model predictions of the differential cross sections for $a_0(980)$
photoproduction in the $\pi^0\eta^0$ channel (left) and $f_0(980)$ 
photoproduction
in the $\pi^0\pi^0$ channel (right).The photon energies are 3.5 (top), 5.0
(centre) and 9.0 (bottom) GeV and the units are $\mu b$ GeV$^{-2}$. The KL 
model predictions are about a factor of 3/4 lower for the
$a_0(980)$ and a factor of 2 lower for the $f_0(980)$.}
\label{a0_f0_edep}
\end{figure}
\end{center}

Predictions of the cross sections for $a_0(980)$ in the $\pi^0\eta^0$ channel
and octet $f_0(980)$ photoproduction in the $\pi^0\pi^0$ channel at $E_\gamma
=$ 3.5, 5.0 and 9.0 GeV are given in Fig. \ref{a0_f0_edep}. The advantage of
the $\pi^0\pi^0$ channel for the $f_0(980)$ is that it automatically excludes
any vector-meson contribution to the final state.

The multiplicity of parameters for the $a_0(980)$ and the $f_0(980)$ lead to
very different $K\bar{K}$ and total widths. In principle $K^+K^-$
photoproduction would provide a check on our model through interference with
the $\phi(1020)$, just as the $f_0(980)$ is seen in $\pi^+\pi^-$
photoproduction through interference with the $\rho(770)$. There are two
relevant experiments \cite{DESY1978,DNPL1982} that provide
indications of a scalar amplitude, albeit with large errors. A
theoretically-based analysis \cite{BLS05} of these data also indicates the
presence of a scalar amplitude, although again with large errors. The
prediction from our model for the $K^+K^-$ scalar cross section lies within
the range of these analyses but their errors are too large to provide any
constraint.

\section{Conclusions}

We have presented a simple parameter-free model for $a_0(980)$ and $f_0(980)$
photoproduction. Within the context of the radiative decay model of
Refs. \cite{CDK02,CDK03} the results imply that the $a_0(980)$ and $f_0(980)$ have
a significant compact $n\bar{n}$ scalar ground
state component in their wave function. This assignment looks quite natural in
naive quark model calculations. For example in Ref. \cite{GI} and
more recently in Ref. \cite{qqbar}, $1^3P_0$ states made of light quarks are
predicted to exist at $1$ GeV, and the $a_0(980)$ and $f_0(980)$ are natural
candidates for such states.

The $K \bar K$ threshold proximity should lead to a
significant admixture of a $K \bar K$ molecular component in the wave function
of scalar resonances, so that a bare $n \bar n$ scalar seed is to be coupled
to pseudoscalar meson pairs. The relative weight of the molecular component
depends on the strength of this coupling, a scenario advocated in
Ref. \cite{baru04}.
Our findings are in favour of such a scenario. Although we cannot quantify the
relative weights of quark and molecular components, we indicate strongly the
presence of the former.

The incompatibility of the $f_0(980)$ couplings to $\pi\pi$ and $K\bar{K}$
between Refs. \cite{kloe08,achasov11,achasov12} on the one hand and
Refs. \cite{martin77,estabrooks79,bugg96,kaminski99,ochs13} on the other is an issue
that clearly needs to be resolved. Photoproduction of $\pi^0\pi^0$ and $K^+K^-$
could achieve this. The minimum experimental requirements are differential
cross sections, plane-polarised beam asymmetries (to obtain information on the
unnatural parity exchange), the resonance line shape (to obtain information on
the background) and sufficiently high energy to eliminate contamination from
$N^*$ production and decay.

\end{document}